\documentclass[]{interact}

\usepackage{epstopdf}
\usepackage[caption=false]{subfig}
\usepackage{fancyhdr}
\usepackage[numbers,sort&compress]{natbib}
\bibpunct[, ]{[}{]}{,}{n}{,}{,}

\usepackage{amsmath}
\usepackage{mathtools}
\usepackage[table]{xcolor}
\usepackage{datetime}
\usepackage{fancyvrb, float}
\usepackage[hidelinks]{hyperref}
\usepackage{booktabs,bm,tabularx}
\usepackage{makecell, caption, booktabs, multirow, siunitx}

\usepackage{lipsum, fancyvrb}
\usepackage{listings}

\theoremstyle{plain}

\theoremstyle{definition}

\theoremstyle{remark}

\begin{document}

\jname{Asian Journal of Statistics and Applications}

\title{Bayesian Joint Modeling of Interrater and Intrarater Reliability with Multilevel Data}

\author{
\name{Nour Hawila\textsuperscript{a} and Arthur Berg\textsuperscript{b}\thanks{CONTACT Author\textsuperscript{b}. Email: berg@psu.edu}}
\affil{\it
\textsuperscript{a}Bristol Meyers Squibb\\
\textsuperscript{b}Division of Biostatistics \& Bioinformatics, Penn State University
}
}

\maketitle

\pagestyle{fancy}
\fancyhead[RO]{\it Hawila and Berg}
\fancyhead[LO]{\it Asian Journal of Statistics and Applications}
\pagenumbering{arabic}

\begin{abstract}
We formulate three generalized Bayesian models for analyzing interrater and intrarater reliability in the presence of multilevel data. Stan implementations of these models provide new estimates of interrater and intrarater reliability. We also derive formulas for calculating marginal correlations under each of the three models. Comparisons of the kappa estimates and marginal correlations across the different models are presented from two real-world datasets. Simulations demonstrate properties of the different measures of agreement under different model assumptions.
\end{abstract}

\begin{keywords}
Reliability; Bayesian; Hierarchical; Nested; Stan
\end{keywords}

\section{Introduction}

It is often important to report on the inconsistent classification of the ratings from different raters (interrater) as well as the inconsistency of the ratings of a single rater's repeated ratings (intrarater). Although often used interchangeably, the primary distinction between reliability and agreement is that agreement is defined as the degree to which classifications are identical, whereas reliability focuses on the extent of variability and error inherent in a measurement \citep{Gisev:13}. Interrater reliability can be defined as the extent to which two or more raters agree in classifying a common observation. It is a helpful measure in assessing the consistency of the implementation of a rating system \citep{Lange:11}. Intrarater reliability refers to the consistency of one rater's classification over multiple time points and is best determined when multiple trials are administered over a short period of time \citep{Sheel:18}. Intrarater reliability is helping in verifying the reproducibility of clinical measurements \citep{Gwet:08}.

Many articles have discussed different measures of association for assessing interrater and intrarater reliability, and are mostly dependent on the type of outcome \citep{Donner:82}. Commonly used measures for the case of binary outcomes include Cohen's Kappa statistic with a corrected standard error and some of its variants \citep{Cohen:60,Fleiss:71,Conger:80}, whereas the intraclass correlation coefficient \citep{Bartko:66} is generally restricted to continuous outcomes. Generalized estimating equations (GEE) are also used for modelling outcomes where Kappa estimates are calculated as a function of the GEE parameters \citep{Williamson:97}. \cite{Gonin:20} also proposed a weighted interrater Kappa calculation based on a GEE approach for categorical data. All of these models utilize a frequentist approach and only estimate interrater reliability and intrarater reliability separately but not jointly.

Joint estimation of interrater and intrarater reliability addresses several key issues inherent in the separate estimation approaches commonly found in the literature. First, by jointly modeling these two types of reliability, we can more accurately capture the interdependencies between the consistency of different raters and the repeatability of the same rater over time. This is particularly important in studies where the same subjects are assessed multiple times by the same raters, as the correlation between assessments can significantly influence the overall reliability of the measurements.

Moreover, joint modeling allows for the simultaneous estimation of both types of reliability, providing a more comprehensive understanding of the measurement process. This is crucial for ensuring the validity of conclusions drawn from the data, especially in clinical settings where decision-making depends heavily on the precision and repeatability of diagnostic assessments. By estimating these reliabilities together, we can also better account for variability between subjects and across time points, leading to more robust estimates of measurement error. Separately estimating inter- and intra-rater reliability ignores the dependencies in the data, whereas joint estimation may provide more accurate reliability estimates, better statistical efficiency, and less bias.

In this paper, we compare three fully Bayesian joint models of interrater and intrarater reliability of increasing complexity. The first model, which we refer to as the Bayesian independent (BIN) model, expands on the model presented in \cite{Nelson:08} by converting it to a Bayesian model and incorporating a random effect for time thus allowing intrarater reliability to also be captured. The second model, which we refer to as the Bayesian partially nested (BPN) model, expands on the model presented in \cite{Nelson:20} by converting it to a Bayesian model and incorporating multiple time points. The third model, which we refer to as the Bayesian fully nested (BFN) model, provides a Stan implementation of the WinBUGS Bayesian model presented in \cite{Hsiao:11} but with more flexible priors.

In Section \ref{sec:methods} we describe both probability-based (Section \ref{sec:prob_based}) and model-based (Section \ref{sec:model_based}) methods in estimating the probability of a positive classification and the resulting interrater and intrarater reliability calculations. Formulas for calculating model-based marginal correlations for each of the three models are provided (Section \ref{sec:marginals}). In Section \ref{sec:sims and examples}, we show the results of applying the proposed Bayesian models to two real-world datasets and provide simulation results aimed at assessing the properties of the different measures of association depending on various model assumptions. We conclude the paper with an overall summary of the findings along with limitations of the proposed methods.

\section{Methods}
\label{sec:methods}
We consider a three-level nested data structure $\{Y_{ijk}\}$ for subject $i$, evaluated by rater $j$, taken at time $k$. This type of data structure is common in observational studies in social and behavioral science, with subjects nested within clusters \citep{Hove:21}. The quality of rating procedures is often of interest with particular emphasis on interrater and intrarater reliability. In the case of continuous outcomes,  the intraclass correlation coefficient (ICC) is typically used to quantify the degree to which different raters resemble each other or the extent to which raters resemble themselves at two different time points. For the scope of this paper, we focus on the case of dichotomous outcomes where $J$ raters evaluate $I$ subjects at $K$ different time points. 

\subsection{Probability-based Methods}
\label{sec:prob_based}
We first consider the simple case of $I$ subjects evaluated by $J$ raters, where interrater reliability is of interest. Denoting the response (classification) of the $j^{th}$ rater on the $i^{th}$ subject by $Y_{ij}$, a disagreement random variable $Z_i$ between raters $j$ and $j'$ evaluating subject $i$ could be constructed. It is assumed that $Z_i$ is common for all pairs $(j,j')$. Kappa coefficients for assessing interrater reliability between raters $j$ and $j'$ are defined by \begin{equation}
    \label{eq:kappa_general}
    \kappa = 1-\frac{E(Z_i)}{\bar E(Z_i)}
\end{equation} 
where $E(Z_i)$ is the expectation of $Z_i$ and $\bar E(Z_i)$ is the expectation of $Z_i$ assuming statistical independence of raters $j$ and $j'$, i.e. $P(Y_{ij}=k_1,Y_{ij'}=k_2)=P(Y_{ij}=k_1)P(Y_{ij'}=k_2)$ \citep{Vanbelle:17}.

\subsubsection{Cohen's Kappa}
\label{sec:cohen}
Cohen's Kappa coefficient in particular is obtained when $Z_i = 1-I(Y_{ij},Y_{ij'})$ where $I(Y_{ij},Y_{ij'})=1$ if $Y_{ij}=Y_{ij'}$ and $0$ otherwise. Assuming equal joint probabilities between raters $j$ and $j'$ across the subjects -- i.e.~$P(Y_{ij}=k_1)=P(Y_{ij'}=k_1)$ for $k_1=0,1$ --  the expectations for Cohen's Kappa are given by
\begin{align}
    E(Z_i) &= P(Y_{ij}=1,Y_{ij'}=0) + P(Y_{ij}=0,Y_{ij'}=1) \\
    \bar E (Z_i) &= P(Y_{ij}=1) P(Y_{ij'}=0) + P(Y_{ij}=0) P(Y_{ij'}=1)
\end{align}
We can now define 
\begin{align}
    p_o &= 1- E(Z_i) = \text{Pr}(Y_{ij}=1,Y_{ij'}=1) + \text{Pr}(Y_{ij}=0,Y_{ij'}=0)\\
    p_c &= 1- \bar E(Z_i) = \text{Pr}(Y_{ij}=1)\text{Pr}(Y_{ij'}=1) + \text{Pr}(Y_{ij}=0)\text{Pr}(Y_{ij'}=0)
\end{align}
to be the observed and expected agreement due to chance respectively.
Cohen's Kappa is more commonly known in following form
\begin{equation}
\label{eq:cohen.c}
    \kappa_c = \frac{p_o-p_c}{1-p_c}.
\end{equation}
Another formulation for $p_c$, without assuming equal joint probabilities between raters is given by
\begin{equation}
    p_c = \text{Pr}(Y_{ij}=1,Y_{i'j'}=1) + \text{Pr}(Y_{ij}=0,Y_{i'j'}=0)
\end{equation}
The true measure of chance agreement $p_c$ is the probability that two randomly selected raters from a population make identical classifications on two different randomly chosen subjects \citep{Nelson:08}.
The definitions of $p_o$ and $p_c$ are left relatively broad because the formulas are adjusted depending on whether interrater reliability or intarrater reliability is of interest and the nesting structure. 
For a three-level nested structure in particular,
\begin{equation}
\label{eq:po}
  p_o =
    \begin{cases}
      \Pr(Y_{ijk} =Y_{ij'k})& \text{for interrater reliability}\\
      \Pr(Y_{ijk} =Y_{ijk'}) & \text{for intrarater reliability}
      \end{cases}       
\end{equation}

\begin{equation}
\label{eq:pc}
  p_c =
    \begin{cases}
      \Pr(Y_{ijk} =Y_{i'j'k})& \text{for interrater reliability}\\
      \Pr(Y_{ijk} =Y_{i'jk'})& \text{for intrarater reliability}
      \end{cases}       
\end{equation}

Like Cohen's Kappa, Scott's Pi is also applicable when evaluating categorical outcomes and therefore also binary data. Scott's Pi is calculated using the formula given in \eqref{eq:cohen.c} but with $p_c$ calculated slightly differently. Scott's Pi assumes the distribution of the raters is the same and therefore $p_c$ is calculated using squared ``joint proportions'' which are squared arithmetic means of the marginal proportions (whereas Cohen's uses squared geometric means of them) \citep{Scott:55}.
Fleiss' Kappa is a generalization of Cohen's Kappa to multiple raters (or time points)  \citep{Fleiss:71}. Congers Kappa which is a correction of Fleiss' Kappa and will be used throughout this paper when computing Kappa \citep{Conger:80}. In the following two subsections, we present closed-form formulas for computing Fleiss' and Conger's Kappa.

\subsubsection{Fleiss' Kappa (1971)}
\label{sec:fleiss}
Let $N$ represent the total number of subjects and $M$ the number of raters per subject. 
Define $n_{ij}$ to be the number of raters assigning subject $i$ to category $j$ where $j=0,1$. We also define $p_{j}$ to be the proportion of all ratings which were assigned to category $j$ and is given by \[
p_{j} = \frac{1}{NM}\sum_{i=1}^Nn_{ij}.
\]
We note here that $\sum_j{n_{ij}}=M$ and $\sum_j{p_{j}}=1$ 

The agreement between the $M$ raters for the $i^{th}$ subject, $P_i$ is the proportion of agreeing pairs out of all $M(M-1)$ possible pairs of assignment given by 
\[
P_i = \frac{1}{M(M-1)}\left( n_{i1}(n_{i1}-1)+n_{i0}(n_{i0}-1)\right)
\] or equivalently can be written as combinations out of all $\binom{M}{2} = \frac{M(M-1)}{2}$ possible pair combinations as
\[
P_i = \frac{1}{\binom{M}{2}}\left[\binom{n_{i1}}{2}+\binom{n_{i0}}{2}\right].
\]

The overall observed extent of agreement can be measured by the mean of $P_i$s given by:
\[
P_o = \frac{1}{N}\sum_{i=1}^NP_i
\]

If the raters made their assignments by chance, the expected mean proportion of agreement would be $(p_jp_{j'} = p_j^2)$ given by
\[
P_c = p_0^2+p_1^2 = \left[\frac{1}{NM}\sum_{i=1}^Nn_{i0}\right]^2 + \left[\frac{1}{NM}\sum_{i=1}^Nn_{i1}\right]^2.
\]

The Kappa coefficient of agreement between the $m$ raters can now be given by
\[
\kappa= \frac{P_o-P_c}{1-P_c}.
\]

\subsubsection{Conger's Kappa (1980)}
\label{sec:congers}
Using the same notation N, M, for $j=0,1$ $P_o$ and $P_c$ are given by:
\[
P_o = \frac{1}{N}\sum_{i=1}^{N}\frac{n_{i1}(n_{i1}-1)+n_{i0}(n_{i0}-1)}{M(M-1)}
\]

\[
P_c = \bar{p}^2_{+0}-s^2_0/M+\bar{p}^2_{+1}-s^2_1/M
\]
where $s^2_k$ is given by 
$s^2_k = \frac{1}{M-1}(p_{j0}-\bar{p}_{+0})^2 + \frac{1}{M-1}(p_{j1}-\bar{p}_{+1})^2$ and represents the variance of the proportion of $p_{j0}$ and $p_{j1}$.
The Kappa coefficient of agreement between the $m$ raters can now be similarly given  as
\[
\kappa= \frac{P_o-P_c}{1-P_c}.
\]

\subsection{Model-based Methods}
\label{sec:model_based}
The measures typically used in \eqref{sec:prob_based} posses some unfavorable traits. Namely, the measures do not incorporate uncertainty about the estimates in addition to the lack of ability to adjust for covariates that could potentially impact the outcomes and most importantly ignore the nested aspect of the data.
In this section, we propose some and generalize other Bayesian models for modelling $p_{ijk}\vcentcolon=\Pr(Y_{ijk}=1)$, the probability the $i^{th}$ subject is classified as a `success' $(Y=1)$ by the $j^{th}$ rater at the $k^{th}$ time point.

\subsubsection{Independent model}
\label{sec:in}
\cite{Nelson:08} proposed a generalized linear mixed model (GLMM) for the case of a single time point given in \eqref{eq:indep}.

\begin{equation}
    \label{eq:indep}
    g(p_{ij}) = \eta  + u_i + v_j, \quad i=1,...,I, j=1,...,J
\end{equation}
$g(.)$ is a link function (usually chosen to be the probit or logit link) and $\eta$ is the intercept term. The random effects terms are given by $u_i$ and $v_j$ for the $i^{th}$ subject and $j^{th}$ rater which are assumed to be independent and normally distributed with mean 0 and variances $\sigma_u^2$ and $\sigma_v^2$, respectively. Under the logit and probit models, estimates of $p_0$ and $p_c$  for a single time point $(k=1)$ are calculated as a function of $\eta,\sigma^2_u,\sigma^2_v$ which are typically replaced by their estimates from the GLMM. This model formulation is useful for evaluating interrater reliability in the presence of a single time point.

We propose the Bayesian model presented in \eqref{eq:indep2} that incorporates an additional random effects terms $w_k$ for the $k^{th}$ time point in addition to a fixed effects term for the $i^{th}$ subject evaluated by the $j^{th}$ rater at the $k^{th}$ time point. 
\begin{equation}
    \label{eq:indep2}
    g(p_{ijk}) = X_{ijk}\beta^\text{I}  + u_i + v_j +w_k, \quad i=1,...,I, j=1,...,J, k=1,...,K
\end{equation}
Positive values of $u_i$ indicate that the $i^{th}$ subject was more likely to be classified as a success amongst raters across time points, positive values of $v_j$ indicate that the $j^{th}$ rater was more likely to classify subjects as success across time points, and positive values of $w_k$ indicate that raters were more likely to classify subjects as success at the $k^{th}$ time point.
The underlying hierarchical Bayesian structure for model \eqref{eq:indep2} is given by
\begin{align}
\label{eq:indep_bayes}
\begin{split}
   u_i &\sim N(0,\sigma_u^2)\\
   v_j &\sim N(0,\sigma_v^2)\\
   w_k &\sim N(0,\sigma_w^2)\\
   \sigma^2_u,\sigma^2_v,\sigma^2_w &\sim IG(\alpha,\gamma)\\
   \beta^\text{I} &\sim  MVN(\mu^\beta_p,\Sigma^\beta_p) \\
\end{split}
\end{align}
Univariate Normal priors are assumed for the random effect terms with 0 mean and independent variance terms that are assumed to have inverse-gamma hyperpriors with parameters $\alpha$ and $\gamma$. The inverse-gamma parameters are typically chosen to be small and equal to reflect non-informative priors. The fixed effects coefficient term $\beta^\text{I}$ is assumed to have a p-variate Normal prior with known mean $\mu^\beta_p$ and unstructured covariance matrix $\Sigma^\beta_p$.

\subsubsection{Fully nested model}
\label{sec:fn}
\cite{Hsiao:11} proposed a Bayesian model able to assess interrater and intrarater reliability for nested binary data structures. The model is given in \eqref{eq:fn}.

\begin{equation}
    \label{eq:fn}
    \text{logit}(p_{ijk}) = X_{ijk}\beta^\text{F} +u_i+v_{ij} +w_{ijk}, \quad i=1,...,I, j=1,...,J, k=1,...,K
\end{equation}
A logit link function was chosen with a flexible fixed effects term $X_{ijk}$ of length $p$ and accompanying fixed effects coefficients vector $\beta^\text{F}$. A first level of independent random effects is imposed at the subject level through $u_i$, followed by $v_{ij}$ that represents a joint subject/rater random effect and finally a subject/rater/time point joint random effect $w_{ijk}$. Similar to the independent model, $u_i$ are assumed to follow an independent univariate Normal distribution with mean $0$ and variance $\sigma_u^2$. The joint subject/rater random effects $v_{i.}$ are assumed to be $J-$variate Normal with mean $\begin{bmatrix} 0 & 0 & ... & 0\end{bmatrix}^T$ and covariance matrix $\Sigma_v^\text{F}=D_v^\text{F}\Omega_v^\text{F} D_v^\text{F}$ where $\Omega_v^\text{F}$ is the correlation matrix with unit entries on the diagonal and $\rho_v^\text{F}$ entries on the off-diagonal, and $D_v^\text{F}=\sqrt{\text{diag}(\Sigma_v^\text{F})}=[\sigma_{v1},...,\sigma_{vJ}]^T$ so that \begin{align}
\label{eq:fn_cov_inter}
\begin{split}
    \text{Var}(v_{ij})&=\sigma_{vj}^2 \\
   \text{Cov}(v_{ij},v_{ij'})&=\rho_v\sigma_{vj}\sigma_{vj'} \text{ for $j=1,2,...,J$ and $j\neq j'$}
\end{split}
\end{align}
We assume a $JK-$variate Normal distribution for $w_{ijk}$ with similar structure to $v_{ij}$ with 0 mean and covariance matrix $\Sigma_w^\text{F}=D_w^\text{F}\Omega_w^\text{F} D_w^\text{F}$ where $\Omega_w^\text{F}$ is the correlation matrix with unit entries on the diagonal and $\rho_w^\text{F}$ entries on the off-diagonal, and $D_w^\text{F}=\sqrt{\text{diag}(\Sigma_w^\text{F})}=[\sigma_{w11},...,\sigma_{wJK}]^T$ so that \begin{align}
\label{eq:fn_cov_intra}
\begin{split}
    \text{Var}(w_{ijk})&=\sigma_{wjk}^2 \\
   \text{Cov}(w_{ijk},w_{ij'k'})&=\rho_w\sigma_{vjk}\sigma_{vj'k'} \text{ for $j=1,...,J, k=1,...,K$ $j\neq j', k\neq k'$}
\end{split}
\end{align}
We note that the covariance matrices $\Sigma_v^\text{F}$ and $\Sigma_w^\text{F}$ are assumed to be \textit{unstructured} for generalizability, but other assumptions could be made. A \textit{common} covariance structure would assume $\sigma_{vj}=\sigma_{vj'}, \forall j \neq j'$  for $\Sigma_v^\text{F}$ and $\sigma_{vjk}=\sigma_{vj'k'}, \forall j \neq j'$ and $k \neq k'$  for $\Sigma_w^\text{F}$. For $\Sigma_w^\text{F}$, we can also assume \textit{separate} covariance structures for different raters (or time points), i.e. $\sigma_{vjk}=\sigma_{vj'k}, \forall j \neq j'$ (or $\sigma_{vjk}=\sigma_{vjk'}, \forall k \neq k'$). 
The full hierarchical Bayesian structure for model \eqref{eq:fn} is given by
\begin{align}
\label{eq:fn_bayes}
\begin{split}
   u_i &\sim N(0,\sigma_u^2)\\
   v_{i.} &\sim MVN(0,\Sigma_v)\\
   w_{i..} &\sim MVN(0,\Sigma_w)\\
   \sigma^2_u, \sigma^2_{vj},\sigma^2_{wjk}&\sim IG(\alpha,\gamma)\\
   \Omega_v^\text{F},\Omega_w^\text{F} &\sim LKJ(\eta)\\
   \beta^\text{F} &\sim  MVN(\mu^{\beta}_p,\Sigma^{\beta}_p) 
\end{split}
\end{align}
 Lewandowski-Kurowicka-Joe (LKJ) prior distributions are assumed for the correlation matrix with tuning parameter $\eta$ to control the strength of the correlations. The fixed effects coefficient term $\beta^{\text{F}}$ is assumed to have a $p$-variate Normal prior with known mean $\mu^\beta_p$ and unstructured covariance matrix $\Sigma^\beta_p$. 
 
 We call this model the fully-nested model due to the hierarchical nature of modelling the random effects. Due to the complex structure of this model, different correlation measurements are of interest, namely, marginal correlations and correlations among random effects. Correlation among rater random effects $\rho^R=\text{Corr}(v_{ij},v_{ij'})$ was used to imply the strength of similarity among the underlying random effect mechanism for raters $j$ and $j'$ and the marginal correlation between these raters (interrater marginal correlation) was calculated as $\text{Corr}^R=\text{Corr}(\text{logit}(p_{ijk}),\text{logit}(p_{ij'k}))$. Similarly, for time points $k$ and $k'$ the correlation among time point random effects is given by $\rho^T=\text{Corr}(w_{ijk},w_{ijk'})$ and the marginal correlation between these time points (intrarater marginal correlation) by $\text{Corr}^T=\text{Corr}(\text{logit}(p_{ijk}),\text{logit}(p_{ijk'}))$. The correlation among rater and time point random effects is used to infer similarity between raters and time points, whereas the marginal correlation measures are used to differentiate between raters or time points as they examine all dependencies involved. 

\subsubsection{Partially nested model}
\label{sec:pn}
Twelve years after \cite{Nelson:08} proposed the independent GLMM, \cite{Nelson:20} proposed an ordinal GLMM model with a crossed random effects structure given in \eqref{eq:pn}. The ordinal GLMM model uses a probit link function to reflect an underlying continuous outcome, and creates a categorized version of the latent variable with $C$ levels, evalued by $J$ raters at $K=2$ time points. The primary usage of this model was to evaluate intrarater reliability across two time points. The model is given by

\begin{equation}
    \label{eq:pn}
    g(p_{ijk}) = \alpha_c - (\beta^\text{P} x_k+u_i+v_{jk}), \quad i=1,...,I, j=1,...,J, k=1,2
\end{equation}
where $\alpha_c$ represents the cutoffs defining the ordinal scaling with $\alpha_0=-\infty$, $\alpha_C=+\infty$ and $ c=1,...,C-1$, $x_k$ is an indicator variable for the $k^{th}$ time point. A fixed effects coefficient $\beta^\text{P}$ provides an overall adjustment for raters at different time points. The subject level random effects $u_i$ are assumed to be independent and normally distributed with mean 0 and variances $\sigma_u^2$, whereas the rater/time point joint random effects $v_{j.}$ are assumed to be bivariate Normal (BVN) with mean $\begin{bmatrix} 0 & 0\end{bmatrix}^T$ and covariance matrix $\Sigma_v^\text{P}=D^\text{P}\Omega^\text{P} D^\text{P}$ where $\Omega^\text{P}$ is the correlation matrix with unit entries on the diagonal and $\rho_v^\text{P}$ entries on the off-diagonal, and $D^\text{P}=\sqrt{\text{diag}(\Sigma_v^\text{P})}=[\sigma_{v1},\sigma_{v2}]^T$ so that \begin{align}
\label{eq:pn_cov_freq}
\begin{split}
    \text{Var}(v_{jk})&=\sigma_{vk}^2 \\
   \text{Cov}(v_{jk},v_{jk'})&=\rho_v\sigma_{vk}\sigma_{vk'} \text{ for $k=1,2$ and $k\neq k'$}
\end{split}
\end{align}

\cite{Nelson:20} also developed integrals for calculating the observed intrarater association and chance intrarater association using the parameters $(\beta, \sigma_u^2,\Sigma_v^\text{P})$ of the model given in \eqref{eq:pn}. To prevent overestimating or underestimating the strength of association between raters, and to account for chance association, \cite{Nelson:20} suggest minimizing the impact of chance association on the Kappa estimate by selecting thresholds $\alpha_{1 \text{min}},...,\alpha_{C-1 \text{min}}$ such that the chance intrarater association integrand is minimized. 
A form of Cohen's Kappa \eqref{eq:cohen.c}, could also be calculated based on model \eqref{eq:pn} by computing the measure for a single rater's paired classification as 
\begin{equation}
    \label{eq:pn_CohensKappa}
    \kappa_c^{\text{avg}} = \frac{p_o^{\text{avg}}-p_c^{\text{avg}}}{1-p_c^{\text{avg}}},
\end{equation}
where $p_o^{\text{avg}}$ and $p_c^{\text{avg}}$ are the average observed and chance interrater (or intrarater) association taken for all pairs of raters $j$ and $j'$ (or time points $k$ and $k'$).

To make this model compatible the case of a binary outcome $(C=2)$, and $K$ time points, an updated version of \eqref{eq:pn} is given by
\begin{equation}
    \label{eq:pn2}
    g(p_{ijk}) = \eta - (\beta_1^{\text{P}} x_k+u_i+v'_{jk}), \quad i=1,...,I, j=1,...,J, k=1,...,K
\end{equation}
where $\beta_1^{\text{P}}$ is a $K\times1$ coefficients vector where the $k^{th}$ entry represents the fixed effect of the $k^{th}$ time point. $v'_{jk}$ is now K-variate Normally distributed with mean $\begin{bmatrix} 0 & 0 & ... & 0\end{bmatrix}^T$ and covariance matrix $\Sigma_v^\text{P'}=D^\text{P'} \Omega^\text{P'} D^\text{P'}$ with similar formulation to $\Sigma_v^\text{P}$ but $K-$dimensional so that 
\begin{align}
\label{eq:pn_cov_bayes}
\begin{split}
    \text{Var}(v_{jk})&=\sigma_{vk}^2 \\
   \text{Cov}(v_{jk},v_{jk'})&=\rho_v\sigma_{vk}\sigma_{vk'} \text{ for $k=1,2,...,K$ and $k\neq k'$}
\end{split}
\end{align}
Incorporating multiple time points allow for a generalized assessment of intrarater reliability.
The underlying hierarchical Bayesian structure for model \eqref{eq:pn2} is given by
\begin{align}
\label{eq:pn_bayes}
\begin{split}
   u_i &\sim N(0,\sigma_u^2)\\
   v_{j.} &\sim MVN(0,\Sigma_v^\text{P'})\\
   \sigma^2_u, \sigma^2_{vk}&\sim IG(\alpha,\gamma)\\
   \Omega^\text{P'} &\sim LKJ(\eta)\\
   \beta_1^{\text{P}} &\sim  MVN(\mu^{\beta}_K,\Sigma^{\beta}_K) 
\end{split}
\end{align}
Univariate Normal priors are assumed for the subject random effect term with 0 mean and independent variance terms that are assumed to have inverse-gamma hyperpriors with parameters $\alpha$ and $\gamma$. The inverse-gamma parameters are typically chosen to be small and equal to reflect non-informative priors. Multivariate Normal priors are assumed for the rater/time point joint random effect with 0 mean and covariance matrix as given in \eqref{eq:pn_cov_bayes}. The fixed effects coefficient term $\beta_1^{\text{P}}$ is assumed to have a p-variate Normal prior with known mean $\mu^\beta_K$ and unstructured covariance matrix $\Sigma^\beta_K$. 

\subsection{Marginal Correlations and Correlations between random effects}
\label{sec:marginals}
After fitting one of the proposed Bayesian models from the previous section, two different measures other than Kappa could be of interest. Namely, the marginal correlation between two raters (or two time points) which is defined as the correlation between the observations made by raters $j$ and $j'$ (or time points $k$ and $k'$). 
\begin{align}
    Corr^R &= Corr(g(p_{ijk}),g(p_{ij'k}))\\
    Corr^T &= Corr(g(p_{ijk}),g(p_{ijk'}))
\end{align}

Another measure of interest is the correlation between random rater effects $\rho^R$ or between random time point effects $\rho^T$. These measures differ based on the type of model selected (independent, partially nested, or fully nested). 

\subsubsection{Independent Model}
For the independent model, the subject, rater and time point random effects are a priori univariate independent Normally distributed. The marginal correlations based on this model are given by:
\begin{align}
     Corr^R_\text{IN} &= \frac{\sigma^2_u + \sigma^2_w}{\sigma^2_u + \sigma^2_v + \sigma^2_w}\\
     Corr^T_\text{IN} &= \frac{\sigma^2_u + \sigma^2_v}{\sigma^2_u + \sigma^2_v + \sigma^2_w}
\end{align}
The correlation between rater random effects and time point random effects are $\rho^R = \rho^T = 0$ due to independence of the random effects. 

\subsubsection{Fully Nested Model}
Assuming a fully nested Bayesian model as given in \ref{eq:fn}, the marginal correlations between two raters and two time points are given by
\begin{align}
     Corr^R_\text{FN} &= \frac{\sigma^2_u + \rho_R\times\sigma^2_v}{\sigma^2_u + \sigma^2_v + \sigma^2_w}\\
     Corr^T_\text{FN} &= \frac{\sigma^2_u + \rho_T \times\sigma^2_w}{\sigma^2_u + \sigma^2_v + \sigma^2_w}
\end{align}
where $\rho_R$ is the correlation between rater random effects and $\rho_T$ is the correlation between time point random effects.
\subsubsection{Partially Nested Model}
Assuming a Bayesian partially nested model as given in \ref{eq:pn2}, the  marginal correlations between two raters and two time points are given by
\begin{align}
     Corr^R_\text{PN} &= \frac{\sigma^2_u}{\sigma^2_u + \sigma^2_{vk}}\\
     Corr^T_\text{PN} &= \frac{\sigma^2_u + \rho_T\sigma_{vk}\sigma_{vk'}}{\sigma^2_u + \sigma^2_{vk}}
\end{align}
The correlation between rater random effects is $\rho^R = 0$ and $\rho^T$ is the correlation between time point random effects. 

\section{Simulation and Examples}
\label{sec:sims and examples}
\subsection{Software and Implementation}
All the developed functions that will allow users to implement the three Bayesian models proposed in Sections \ref{sec:in}, \ref{sec:fn} and \ref{sec:pn} using RStan \citep{Stan:20} in addition to easily computing interrater and intrarater reliability posterior estimates, credible intervals and simulation results are available through \url{https://github.com/NourHawila/IRR}.

\subsection{Data Examples}
\label{sec:examples}

All three Bayesian models -- BIN, BPN and BFN -- are fit to two real-world datasets. The resulting model fits are subsequently used to explore model properties through simulations. 

\subsubsection{Dataset: Running gait}
\label{sec:drone}
In a crossover study published in 2022, \cite{Lafferty:20} recruited 32 cross-country, track and field, and recreational athletes with current running mileage of at least 15 km per week to compare indoor and outdoor running environments. Athletes first ran on a treadmill in an indoor environment recorded using static video, followed by outdoor video recorded by a drone. Three judges were shown both videos, two weeks apart, and independently performed running gait analysis on each foot for each of the 32 runners, with the goal of assessing interrater and intrarater reliability in addition to indoor vs. outdoor agreement. This dataset consists of a total of $768$ data points (32 runners, 3 raters, 2 time points, 2 feet, 2 locations) in a complete block design. Fleiss' Kappa was used to assess reliability in \cite{Lafferty:20} after averaging across the nested structures.

\subsubsection{Dataset: Radiograph}
\label{sec:radiograph}
The quality of root canal is typically evaluated by a radiographic assessment that is known to be necessary yet subjective. A study was conducted to assess the consistency and accuracy of the radiographic evaluation of 7 endodontists who partook in a training course and evaluated radiographs from 35 participants before and after the training \citep{Chueh:03}. The quality of endodontic treatment is defined as ‘‘good’’ if the filling reaches an adequate length within 2 mm of the radiographic apex and if the complete obturation is in the apical one-third of the root canal \citep{Hsiao:11}. This dataset consists of a total of $490$ data points (35 subjects, 7 raters, 2 time points) in a complete block design. \cite{Hsiao:11} used the BFN model to assess interrater and intrarater reliability using WinBUGS.

\subsubsection{Model fits of two real-world datasets}

The three Bayesian models were fit to each of the two datasets. Each of the models used 2000 iterations, 200 warmup iterations per chain across 2 chains. The functions \texttt{model\_BIN}, \texttt{model\_BPN}, and \texttt{model\_BFN}, available at \url{https://github.com/NourHawila/IRR}, are used to fit the datasets to the respective Bayesian models. The syntax for fitting the three models is presented below. The \texttt{fixed\_eff\_intercept} argument controls whether the fixed effects design matrix includes an intercept term or not. The parameters \texttt{beta\_a} and \texttt{beta\_b} are the hyperpriors for the correlation terms \texttt{rho\_R} and \texttt{rho\_T} which are assumed to follow Beta distributions. The parameters \texttt{gamma\_a} and \texttt{gamma\_b} are the hyperpriors for the variance terms \texttt{sigma\_S}, \texttt{sigma\_R}, and \texttt{sigma\_T} which are assumed to follow inverse gamma distributions. The parameters \texttt{rho\_R\_eta} and \texttt{rho\_T\_eta} are the hyperior shape parameters for the correlation terms \texttt{rho\_R} and \texttt{rho\_T} respectively which are assumed to follow an LKJ correlation distribution.

\begin{description}
    \item[BIN] \texttt{model\_BIN(df, fixed\_eff\_intercept = TRUE, 
beta\_a = 5, beta\_b = 5, gamma\_a = 3, gamma\_b = 1.5, 
beta\_mean =0, beta\_sigma = 1/0.3, niters = 2000, 
nwarmup = 200, nchains = 2)}

\item[BPN] \texttt{model\_BPN(df, cov\_T\_str = "common",  
                              fixed\_eff\_intercept = TRUE, beta\_a = 5, beta\_b = 5, gamma\_a = 3, gamma\_b = 1.5, beta\_mean = 0, beta\_sigma = 1/0.3, betak\_mean = 0, betak\_sigma = 1/0.3, rho\_T\_eta = 1, niters = 2000, nwarmup = 200, nchains = 2) )}

\item[BFN] \texttt{model\_BFN(df, cov\_R\_str = "common",cov\_T\_str = "common", fixed\_eff\_intercept = TRUE, beta\_a = 5, beta\_b = 5, gamma\_a = 3, gamma\_b = 1.5, beta\_mean = 0, beta\_sigma = 1/0.3, rho\_R\_eta = 1, rho\_T\_eta = 1, niters = 2000, nwarmup = 200, nchains = 2) )}
\end{description}

Table \ref{tab:datafits} presents fits of the three Bayesian models to the two datasets. Based on the leave-one-out information criterion (LOOIC), the best model for the running gait dataset is the BPN model with the BIN model in a close second place. The LOOIC selected the BFN as the best model for the radiograph data, while the other two models displayed substantially larger LOOICs. 

Estimates of the interrater and intrarater Kappa measures based on direct frequentist estimation and model-based methods are presented in Table \ref{tab:mydata}. Using the generated quantities block in Rstan, we simulate data from the posterior predictive distribution and calculate the interrater and intrarater kappa estimates. For the running gait dataset, the LOOIC-selected BPN model produced interrater and intrarater kappa estimates (0.24 and 0.30, respectively) that substantially differ from the frequentist-based estimates (0.44 and 0.45, respectively). Similarly, for the radiograph dataset, the LOOIC-selected BFN model produced interrater and intrarater kappa estimates (0.07 and 0.33, respectively) that are even more different from the frequentist-based estimates (0.40 and 0.72, respectively). These results show that datasets with small number of raters and timepoints can lead to widely varying estimates of interrater and intrarater reliability depending on what approach is used. In the next section, we simulate datasets from different models to evaluate these differences in interrater and intrarater reliability estimation.

\begin{table}[H]
    \centering
    \caption{Posterior mean estimates, effective number of parameters (p\_LOO), and the LOOIC for fitting the three Bayesian models to the two datasets described in Section \ref{sec:examples}}
    \label{tab:datafits}
    \begin{tabular}{|c|c|c|c|}
\hline
\textbf{Model}                 & \textbf{Parms}        & \textbf{Running Gait} & \textbf{Radiograph} \\ \hline
\multirow{9}{*}{\textbf{BIN}}  & $\sigma_S$            & 0.91                  & 1.31                \\ \cline{2-4} 
                               & $\sigma_R$            & 0.79                  & 0.77                \\ \cline{2-4} 
                               & $\sigma_T$            & 0.79                  & 0.78                \\ \cline{2-4} 
                               & $Corr^\text{R}$       & 0.70                  & 0.79                \\ \cline{2-4} 
                               & $Corr^\text{T}$       & 0.70                  & 0.78                \\ \cline{2-4} 
                               & \texttt{p\_LOO}                & 32.02                 & 32.29               \\ \cline{2-4} 
                               & \texttt{LOOIC}                 & 720.89                & 396.54              \\ \hline
\multirow{8}{*}{\textbf{BPN}}  & $\sigma_S$            & 0.89                  & 1.31                \\ \cline{2-4} 
                               & $\sigma_T$            & 0.75                  & 0.70                \\ \cline{2-4} 
                               & $Corr^\text{T}$       & 0.80                  & 0.88                \\ \cline{2-4} 
                               & $\rho^\text{T}$       & 0.50                  & 0.51                \\ \cline{2-4} 
                               & \texttt{p\_LOO}                & 33.17                 & 35.9                \\ \cline{2-4} 
                               & \texttt{LOOIC}                 & \textbf{720.44}       & 398.81              \\ \hline
\multirow{11}{*}{\textbf{BFN}} & $\sigma_S$            & 0.70                  & 0.63                \\ \cline{2-4} 
                               & $\sigma_R$            & 0.65                  & 0.88                \\ \cline{2-4} 
                               & $\sigma_T$            & 0.64                  & 1.47                \\ \cline{2-4} 
                               & $Corr^\text{R}$       & 0.55                  & 0.24                \\ \cline{2-4} 
                               & $Corr^\text{T}$       & 0.84                  & 0.72                \\ \cline{2-4} 
                               & $\rho^\text{R}$       & 0.54                  & 0.42                \\ \cline{2-4}  
                               & $\rho^\text{T}$       & 0.47                  & 0.51                \\ \cline{2-4} 
                               & \texttt{p\_LOO}                & 72.38                 & 82.83               \\ \cline{2-4} 
                               & \texttt{LOOIC}                 & 741.02                & \textbf{318.02}     \\ \hline
\end{tabular}
\end{table}

\begin{table}[H]
    \centering
    \caption{Interrater and intrarater Conger Kappa estimates for the two datasets}
    \label{tab:mydata}
    \begin{tabular}{|c|c|c|c|}
\hline
                               &                       & \textbf{\begin{tabular}[c]{@{}c@{}}Running\\ Gait\end{tabular}} & \textbf{Radiograph} \\ \hline
\multirow{2}{*}{\textbf{Freq}} & $\kappa_\text{inter}$ & 0.44                                                            & 0.40                \\ \cline{2-4} 
                               & $\kappa_\text{intra}$ & 0.45                                                            & 0.72                \\ \hline
\multirow{2}{*}{\textbf{BIN}}  & $\kappa_\text{inter}$ & 0.26                                                            & 0.39                \\ \cline{2-4} 
                               & $\kappa_\text{intra}$ & 0.27                                                            & 0.40                \\ \hline
\multirow{2}{*}{\textbf{BPN}}  & $\kappa_\text{inter}$ & 0.24                                                            & 0.37                \\ \cline{2-4} 
                               & $\kappa_\text{intra}$ & 0.30                                                            & 0.42                \\ \hline
\multirow{2}{*}{\textbf{BFN}}  & $\kappa_\text{inter}$ & 0.18                                                            & 0.07                \\ \cline{2-4} 
                               & $\kappa_\text{intra}$ & 0.26                                                            & 0.33                \\ \hline
\end{tabular}
\end{table}

\subsection{Simulations}
Datasets are simulated from the three different models -- independent (IN), partially-nested (PN), and fully-nested (FN) -- as presented in Section \ref{sec:model_based}. The parameter values used in these model-based simulations are the posterior estimates from the respective model fits presented in Table \ref{tab:datafits}. In this way, the simulated datasets are based on the two real world datasets, and these respective simulations are labeled as `running gait' and `radiograph'. For each dataset-based simulation (running gait and radiograph), and for each model (IN, PN, FN), a total of $148$ datasets were generated. For each of the generated datasets, the three Bayesian models (BIN, BPN, BFN) were fit to the data and their respective LOOIC was recorded. The model with the smallest LOOIC is referred to as the `LOOIC-selected' Bayesian model for that simulation. Over each set of $148$ simulations, the relative proportions of the LOOIC-selected models are presented in Table \ref{tab:LOOmodelfits}. 

When the true simulation is IN, the BIN model is the most common LOOIC-selected model for both dataset references. Similarly, when the true simulation is PN, the BPN model is the most common LOOIC-selected model for both dataset references. However, when the the true simulation is FN, the BFN model is the most common LOOIC-selected model for the running gait dataset reference but the BPN model is the most common LOOIC-selected model for the radiograph dataset reference.

\begin{table}[H]
\centering
\caption{Relative proportions of LOOIC-selected models for $148$ simulations with each row adding up to $100\%$ and the most commonly selected model for each set of simulations is bolded}
\label{tab:LOOmodelfits}
\begin{tabular}{|c|c|l|l|l|}
\hline
\textbf{\begin{tabular}[c]{@{}c@{}}Dataset\\ Reference\end{tabular}}                                        & \textbf{\begin{tabular}[c]{@{}c@{}}Sim\\ Model\end{tabular}} & \multicolumn{1}{c|}{\textbf{BIN}} & \multicolumn{1}{c|}{\textbf{BPN}} & \multicolumn{1}{c|}{\textbf{BFN}} \\ \hline
\rowcolor[HTML]{FFFFFF} 
\cellcolor[HTML]{FFFFFF}                                                                                  & IN                                                                & \textbf{77.7\%}                              & 18.2\%                              & 4.1\%                               \\ \cline{2-5} 
\rowcolor[HTML]{FFFFFF} 
\cellcolor[HTML]{FFFFFF}                                                                                  & PN                                                                & 14.2\%                              & \textbf{85.1\%}                              & 0.7\%                               \\ \cline{2-5} 
\rowcolor[HTML]{FFFFFF} 
\multirow{-3}{*}{\cellcolor[HTML]{FFFFFF}\textbf{\begin{tabular}[c]{@{}c@{}}Running\\ Gait\end{tabular}}} & FN                                                                & 7.4\%                               & 37.8\%                              & \textbf{54.7\%}                            \\ \hline
\rowcolor[HTML]{FFFFFF} 
\cellcolor[HTML]{FFFFFF}                                                                                  & IN                                                                & \textbf{73.6\%}                             & 21.6\%                              & 4.7\%                               \\ \cline{2-5} 
\rowcolor[HTML]{FFFFFF} 
\cellcolor[HTML]{FFFFFF}                                                                                  & PN                                                                & 12.2\%                              & \textbf{82.4\%}                              & 5.4\%                               \\ \cline{2-5} 
\rowcolor[HTML]{FFFFFF} 
\multirow{-3}{*}{\cellcolor[HTML]{FFFFFF}\textbf{\begin{tabular}[c]{@{}c@{}}Radio-\\ graph\end{tabular}}} & FN                                                                & 1.4\%                               & \textbf{98.6\%}                            & 0\%                                 \\ \hline
\end{tabular}
\end{table}

For each dataset-based simulation (running gait and radiograph), and for each model (IN, PN, FN), approximate theoretical interrater and intrarater reliability measures are generated by simulating 10,000 datasets and taking the average kappa estimates. The resulting `true' kappa parameters are displayed in Table \ref{tab:sim_kappa}. The frequentist-based and model-based (BIN, BPN, BFN, LOOIC-selected) interrater and intrarater kappa estimates are calculated for each of the $148$ simulated datasets and the average parameter estimates and root mean square error (RMSE) performances are presented. The overall performance of each method (frequentist, BIN, BPN, BFN, and LOOIC-selected) over all $148 \times 3 = 444$ simulated datasets within each dataset reference is highlighted in blue. The Bayesian model consistent with the respective data simulation is highlighted in yellow. The method with the smallest RMSE is bolded in each row of the table.

The simulations show all three Bayesian models perform similarly in terms of interrater and intrarater estimation. However, the frequentist-based estimation of interrater and intrarater reliability is shown to be worse than each of the Bayesian models for every simulation scenario. All three Bayesian models tend to perform similarly with regard to interrater and intrarater reliability estimation; the Bayesian model consistent with the data simulation model was not always optimal though it is consistently close to the optimal model.

\begin{table}[H]
    \centering
    \caption{Kappa simulation results showing mean estimates (RMSE)}
    \label{tab:sim_kappa}
    \resizebox{\columnwidth}{!}{\begin{tabular}{|c|c|c|c|c|c|c|c|c|}
\hline
\textbf{\begin{tabular}[c]{@{}c@{}}Model\\ Parms\end{tabular}}                    & \textbf{Sim}                                                         & \textbf{$\bm{\kappa}$} & \textbf{True}        & \textbf{Freq}                 & \textbf{BIN}                                   & \textbf{BPN}                          & \textbf{BFN}                                   & \textbf{LOO}                           \\ \hline
                                                                                  &                                                                      & inter                  & 0.261                & 0.265 (0.111)                 & \cellcolor[HTML]{FFFFC7}0.266 (0.106)          & 0.266 (0.106)                         & \textbf{0.254 (0.102)}                         & 0.265 (0.105)                          \\
                                                                                  & \multirow{-2}{*}{IN}                                                 & intra                  & 0.271                & 0.265 (0.115)                 & \cellcolor[HTML]{FFFFC7}0.268 (0.108)          & 0.263 (0.107)                         & \textbf{0.271 (0.103)}                         & 0.267 (0.108)                          \\ \cline{2-9} 
                                                                                  &                                                                      & inter                  & 0.236                & 0.231 (0.06)                  & 0.242 (0.056)                                  & \cellcolor[HTML]{FFFFC7}0.233 (0.056) & \textbf{0.226 (0.055)}                         & 0.233 (0.057)                          \\
                                                                                  & \multirow{-2}{*}{PN}                                                 & intra                  & 0.297                & 0.297 (0.086)                 & 0.321 (0.068)                                  & \cellcolor[HTML]{FFFFC7}0.299 (0.071) & \textbf{0.325 (0.065)}                         & 0.3 (0.071)                            \\ \cline{2-9} 
                                                                                  &                                                                      & inter                  & 0.178                & 0.174 (0.066)                 & 0.201 (0.063)                                  & 0.194 (0.061)                         & \cellcolor[HTML]{FFFFC7}\textbf{0.177 (0.056)} & 0.182 (0.058)                          \\
                                                                                  & \multirow{-2}{*}{FN}                                                 & intra                  & 0.259                & 0.261 (0.082)                 & \textbf{0.262 (0.065)}                         & 0.246 (0.07)                          & \cellcolor[HTML]{FFFFC7}0.286 (0.067)          & 0.272 (0.068)                          \\ \cline{2-9} 
                                                                                  &                                                                      & inter                  &                      & \cellcolor[HTML]{DAE8FC}0.082 & \cellcolor[HTML]{DAE8FC}0.078                  & \cellcolor[HTML]{DAE8FC}0.078         & \cellcolor[HTML]{DAE8FC}\textbf{0.074}         & \cellcolor[HTML]{DAE8FC}0.077          \\
\multirow{-8}{*}{\textbf{\begin{tabular}[c]{@{}c@{}}Running\\ Gait\end{tabular}}} & \multirow{-2}{*}{\begin{tabular}[c]{@{}c@{}}Avg\\ RMSE\end{tabular}} & intra                  & \multirow{-2}{*}{--} & \cellcolor[HTML]{DAE8FC}0.096 & \cellcolor[HTML]{DAE8FC}0.083                  & \cellcolor[HTML]{DAE8FC}0.085         & \cellcolor[HTML]{DAE8FC}\textbf{0.08}          & \cellcolor[HTML]{DAE8FC}0.084          \\ \hline
                                                                                  &                                                                      & inter                  & 0.385                & 0.387 (0.088)                 & \cellcolor[HTML]{FFFFC7}\textbf{0.381 (0.083)} & 0.381 (0.084)                         & 0.371 (0.085)                                  & 0.38 (0.084)                           \\
                                                                                  & \multirow{-2}{*}{IN}                                                 & intra                  & 0.399                & 0.39 (0.113)                  & \cellcolor[HTML]{FFFFC7}\textbf{0.386 (0.105)} & 0.372 (0.106)                         & 0.395 (0.107)                                  & 0.385 (0.107)                          \\ \cline{2-9} 
                                                                                  &                                                                      & inter                  & 0.373                & 0.378 (0.072)                 & 0.373 (0.067)                                  & \cellcolor[HTML]{FFFFC7}0.372 (0.067) & 0.365 (0.069)                                  & \textbf{0.371 (0.067)}                 \\
                                                                                  & \multirow{-2}{*}{PN}                                                 & intra                  & 0.423                & 0.426 (0.075)                 & \textbf{0.439 (0.061)}                         & \cellcolor[HTML]{FFFFC7}0.415 (0.062) & 0.455 (0.064)                                  & 0.421 (0.064)                          \\ \cline{2-9} 
                                                                                  &                                                                      & inter                  & 0.074                & 0.074 (0.052)                 & 0.092 (0.049)                                  & 0.076 (0.049)                         & \cellcolor[HTML]{FFFFC7}\textbf{0.085 (0.049)} & 0.076 (0.049)                          \\
                                                                                  & \multirow{-2}{*}{FN}                                                 & intra                  & 0.331                & 0.327 (0.174)                 & 0.413 (0.147)                                  & 0.308 (0.164)                         & \cellcolor[HTML]{FFFFC7}\textbf{0.341 (0.123)} & 0.309 (0.164)                          \\ \cline{2-9} 
                                                                                  &                                                                      & inter                  &                      & \cellcolor[HTML]{DAE8FC}0.072 & \cellcolor[HTML]{DAE8FC}0.068                  & \cellcolor[HTML]{DAE8FC}0.068         & \cellcolor[HTML]{DAE8FC}0.069                  & \cellcolor[HTML]{DAE8FC}\textbf{0.068} \\
\multirow{-8}{*}{\textbf{\begin{tabular}[c]{@{}c@{}}Radio-\\ graph\end{tabular}}} & \multirow{-2}{*}{\begin{tabular}[c]{@{}c@{}}Avg\\ RMSE\end{tabular}} & intra                  & \multirow{-2}{*}{--} & \cellcolor[HTML]{DAE8FC}0.127 & \cellcolor[HTML]{DAE8FC}0.11                   & \cellcolor[HTML]{DAE8FC}0.118         & \cellcolor[HTML]{DAE8FC}\textbf{0.101}         & \cellcolor[HTML]{DAE8FC}0.119          \\ \hline
\end{tabular}}
\end{table}

\section{Discussion}
\label{sec:discussion}
In this paper we have presented three Bayesian GLMMs and assessed the properties of different model assumptions on the estimation of interrater and intrarater reliability. The proposed models are comprehensive in terms of jointly adjusting for subject, rater and time point effects. The models are also flexible and able to incorporate different prior beliefs and knowledge about specific model parameters through Bayesian modelling. Contrary to commonly used models, the ones presented are not restricted to two raters or time points but are applicable for the general case of $I$ subjects, $J$ raters and $K$ time points.

Some drawbacks in utilizing these methods for assessing interrater and intrarater reliability include the complexity in analyzing all the different parameters of interest and ensuring proper convergence of the model. Selecting the most appropriate model among the three Bayesian models requires a good understanding of the data, though the simulations show that utilizing the LOOIC model selection method will often select the optimal or near-optimal model. As in any Bayesian model fit, the choice of priors may have substantial impact on the posterior estimates, so a sensitivity analysis that varies the priors should be performed.

\bibliographystyle{agsm}
\bibliography{irr}

\end{document}